# Geometrical Approach to Central Molecular Chirality: A Chirality Selection Rule

SALVATORE CAPOZZIELLO[a] AND ALESSANDRA LATTANZI[b]*

[a]*Dipartimento di Fisica "E. R. Caianiello", INFN sez. di Napoli* and [b]*Dipartimento di Chimica,
Università di Salerno, Via S. Allende, 84081, Baronissi, Salerno, Italy*
E-mail: lattanzi@unisa.it

*ABSTRACT*   Chirality is of primary importance in many areas of chemistry and such a topic has been extensively investigated since its discovery. We introduce here the description of central chirality for tetrahedral molecules using a geometrical approach based on complex numbers. According to this representation, it is possible to define, for a molecule having *n* chiral centres, an "index of chirality $\chi$". Consequently, a "chirality selection rule" has been derived which allows to characterize a molecule as achiral, enantiomer or diastereoisomer.

Molecular chirality has a central role in organic chemistry. Most of the molecules of interest to the organic chemist are chiral, so, in the course of the years, a great amount of experimental work has been devoted to the selective formation of molecules with a given chirality, the so called asymmetric synthesis.[1]

Chirality was defined by Lord Kelvin[2] almost one century ago as follows: "I call any geometrical figure, or groups of points, chiral, and say it has chirality, if its image in a plane mirror, ideally realized, cannot be brought to coincide with itself". The corresponding definition of an achiral molecule can be expressed as: "if a structure and its mirror image are superimposable by rotation or any motion other than bond making and breaking, than they are identical". Chiral molecules having *central chirality* contain stereogenic centres.[3] Given two molecules with identical chemical formulas, if they are not superimposable, they are called *enantiomers*. In general, the term chirality has a broader sense, for example, chirality can be due to a spatial isomerism resulting from the lack of free rotation around single or double bonds (which means that the molecule has a *chiral axis*) such as respectively in biphenyl[4] and in allene[5] derivatives, rather than due to the presence of stereogenic centres (Fig. 1).

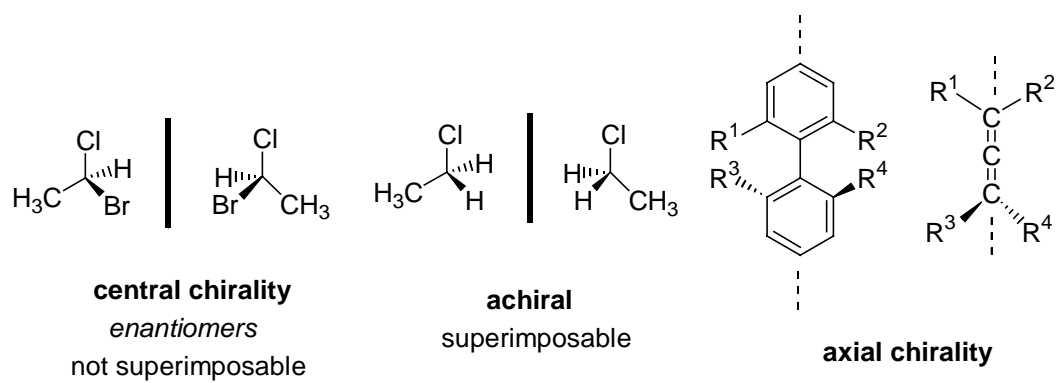

*Fig. 1. Examples of different forms of chirality*

In this contribution, we refer to the term of chirality in the narrow sense of central chirality for the tetrahedral molecules which are the most common class of chiral molecules. When the molecule contains more than one chiral centre, a further definition has to be introduced. In this case, two molecules with identical structural formulas, which are not mirror images of each other and not superimposable, are termed *diastereoisomers* (Fig. 2).

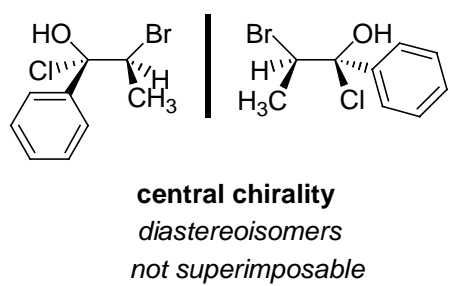

*Fig. 2. Example of a couple of diastereoisomers*

Most properties of molecules are invariant to reflection (scalar properties), when examined in an achiral environment and enantiomers will be identical in many respects such as solubility, density, melting point, chromatographic retention times, spectroscopic behaviour. It is only with respect to properties that change sign, but not magnitude, upon reflection (pseudoscalar properties) such as optical rotation,[6] optical rotatory dispersion (ORD),[7] circular dichroism (CD),[8] vibrational circular dichroism (VCD)[9] that enantiomers differ. In contrast, diastereoisomers exhibit different chemical and physical properties. It is evident that molecular chirality is fundamentally connected to spatial symmetry operations[10] and has the features of a geometrical property. Interestingly, chirality has been recently treated as a continuous phenomenon[11] of achiral symmetry breaking and this approach has brought to the description of molecules as "more or less chiral" just as doors are more or less open. In the present article, on the basis of the dichotomous character of chirality, we describe this topic using a geometrical approach leading us to enucleate algebraic structures[12] of tetrahedral central chirality.

## THEORY AND DISCUSSION

The spatial properties of achiral molecules, enantiomers and diastereoisomers can be considered under the same standard of a geometrical description and we wonder whether some features exist in order to describe such classes of molecules by the same parameters.

Our approach is based on complex numbers since this is a straightforward way to represent the "length" of the bond with respect to the stereogenic centre and the "angular position" with respect to the other bonds. In general, given a tetrahedral molecule with a stereogenic centre, we can always project it on a plane containing the stereogenic centre as in Fig. 3.

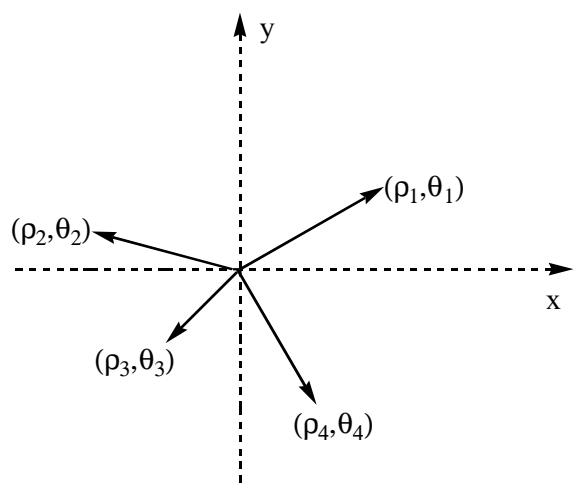

Fig. 3. **Projection of a tetrahedral molecule on a plane containing the stereogenic centre**

Every bond, in the plane {x,y}, can be given in polar representation by

$$\Psi_j = \rho_j e^{i\theta_j} \qquad [1]$$

where $\rho_j$ is the "modulus", i.e. the projected length of the bond, $\theta_j$ is the "anomaly", i.e. the position of the bond with respect to the x, y axes (and then with respect to the other bonds) having chosen a rotation versus. The number $i = \sqrt{-1}$ is the imaginary unit. A molecule with one stereogenic centre is then given by the sum vector

$$\mathcal{M} = \sum_{j=1}^{4} \rho_j e^{i\theta_j} \qquad [2]$$

in any symmetry plane. If the molecule has $n$ stereogenic centres, we can define $n$ planes of projection (one for each centre). Such planes can be parallel among them, but this feature is not essential. Now, the bonds of the different groups bound to the same chiral centre and the bonds among the centres have to be taken into account. So if a molecule with one centre has four bonds, a molecule with two centres has seven bonds and so on. The general rule is

$$n = \text{centres} \quad \Leftrightarrow \quad 4n - (n-1) = 3n+1 \text{ bonds} \qquad [3]$$

assuming, as standard, simply connected tetrahedrons. When atoms are present, which act as "spacers" between the tetrahedral chiral centres, the number of bonds changes from $3n+1$ to $4n$ (as for ex: (1-*sec*-butoxy-butyl)-benzene), but the following considerations for consecutive connected tetrahedrons remain valid. A molecule with $n$ stereogenic centres is then given by the sum vector

$$\mathcal{M}_n = \sum_{k=1}^{n} \sum_{j=1}^{3n+1} \rho_{jk} e^{i\theta_{jk}} \qquad [4]$$

where $k$ is the "centre-index" and $j$ is the "bond-index". Again, for any $k$, a projective plane of symmetry is defined. The couple of numbers $\{\rho, \theta\} \equiv \{0,0\}$ gives the centre in any plane. In other words, a molecule $\mathcal{M}_n$ is assigned by the two sets of numbers

$$\{\rho_{1k}, \ldots \rho_{jk}, \ldots \rho_{(3n+1)k}\}$$

$$[5]$$

$$\{\theta_{1k}, \ldots \theta_{jk}, \ldots \theta_{(3n+1)k}\}$$

where $k$ is the generic tetrahedron (i.e. the centre index). The case $n=1$ gives the simple tetrahedron.

Having in mind the definition of chirality, the behaviour of the molecule under rotation and superimposition has to be studied in order to see if the structure and its mirror image are superimposable. Chirality emerges when two molecules with identical structural formulas are not superimposable. Considering the geometrical representation reported in Fig. 3, a possible situation is the following: let us take into account a rotation of 180° in the space around a generic axis *L* passing through the stereogenic centre. Such an axis can coincide, for the sake of simplicity with one of the bonds. After the rotation two bonds result surimposable while the other two are inverted. The situation can be illustrated by the projection on the plane {x,y} as shown in Fig.4.

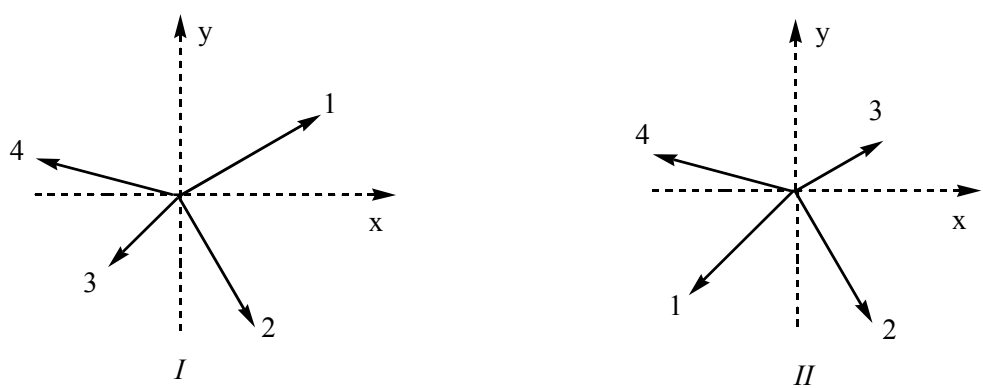

**Fig. 4.** Picture of the projected situations before and after the rotation of a chiral tetrahedron over its mirror image. Groups 2 and 4 coincide while 3 and 1 are inverted.

In formulae, for the inverted bonds, we have

$$\{\Psi_1 = \rho_1 e^{i\theta_1}, \Psi_3 = \rho_3 e^{i\theta_3}\} \Rightarrow \{\overline{\Psi}_1 = \rho_1 e^{i\theta_3}, \overline{\Psi}_3 = \rho_3 e^{i\theta_1}\} \qquad [6]$$

In order to observe the reflection, the four groups must be of different nature (see Fig. 1). This simple observation shows that the chirality is connected with an inversion of two bonds in the projective symmetry plane. On the contrary, if after the rotation and superimposition (Fig. 4), molecule *I* is identical to molecule *II*, we are in an achiral situation. Such a treatment can be repeated for any projective symmetry plane which can be defined for the *n* centres. The possible results are that the molecule is fully invariant after rotation(s) and superimposition with respect to its mirror image (achiral); the molecule is partially invariant after rotation(s) and superimposition, i.e. some tetrahedrons are superimposable while others are not (diastereoisomers); the molecule presents an inversion for each stereogenic centre (enantiomers). The following rule can be derived: central chirality is assigned by the number $\chi$ given by the couple $n, p$ that is

$$\chi = \{n, p\} \tag{7}$$

where

$$\chi = \text{chirality index}$$

$$n = \text{principal chiral number}$$

$$p = \text{secondary chiral number}$$

$n$ is the number of stereogenic centres, $p$ is the number of permutations (at most one for any centre). The constraint

$$0 \leq p \leq n \tag{8}$$

has to hold.

The sequence between achiral and chiral molecules is given by

$$\chi \equiv \{n, 0\} \quad \textit{achiral molecules}$$

$$\chi \equiv \{n, p < n\} \quad \textit{diastereoisomers}$$

$$\chi \equiv \{n, n\} \quad \textit{enantiomers}$$

This definition of chirality is related to the structure of the molecule and its properties under rotations and superimposition.

Such an argument can be re-expressed in an algebraic formulation. Let $\hat{\chi}$ be the central chirality operator acting on a tetrahedron which is given as a column vector $\mathcal{M}$. We have

$$\hat{\chi} \mathcal{M} = \overline{\mathcal{M}} \tag{9}$$

or explicitly

$$\hat{\chi} \begin{pmatrix} \Psi_1 \\ \Psi_2 \\ \Psi_3 \\ \Psi_4 \end{pmatrix} = \begin{pmatrix} \Psi_1 \\ \Psi_2 \\ \Psi_4 \\ \Psi_3 \end{pmatrix} \tag{10}$$

where the groups $\Psi_3$ and $\Psi_4$ result inverted under the action of $\hat{\chi}$.

A possible matrix representation is

$$\hat{\chi} = \begin{pmatrix} 1 & 0 & | & 0 & 0 \\ 0 & 1 & | & 0 & 0 \\ \overline{0} & \overline{0} & | & \overline{0} & \overline{1} \\ 0 & 0 & | & 1 & 0 \end{pmatrix} \qquad [11]$$

It is easy to see that the matrix [11] has a unitary part (upper part) and an anti-unitary part (lower part). It can be considered a composition of a rotation and a reflection. If two of the components $\Psi_j$ are the same chemical group, i.e. $\Psi_3 = \Psi_4$, it follows that $\mathcal{M} \equiv \overline{\mathcal{M}}$ and we have an achiral situation. However, $\hat{\chi}$ can act on any couple of groups $\Psi_j$ and $\Psi_i$ so that, in general, we can define an algebra of $\hat{\chi}_k$ operators[13] where $k$ is an index which runs over the possible permutations.

The eigenvalue problem is expressed by the determinant

$$\det \| \hat{\chi} - \lambda \hat{\mathbf{I}} \| = 0 \qquad [12]$$

where $\lambda$ are the eigenvalues and $\hat{\mathbf{I}}$ is the identity matrix. In our specific case, Eq. [12] reduces to

$$(1-\lambda)^3 (\lambda+1) = 0 \qquad [13]$$

The eigenvalues are $\lambda = \pm 1$, with $\lambda = 1$ triply degenerate. The eigenvector equation for $\hat{\chi}$ can be written as

$$\hat{\chi}\varphi = \lambda\varphi \qquad [14]$$

where $\varphi$ is an eigenvector.

If $\lambda = 1$, the action of $\hat{\chi}$ on the (molecule) eigenvector $\varphi$ gives the "same" vector (achirality);

if $\lambda = -1$, the action of $\hat{\chi}$ on $\varphi$ gives an inversion (chirality), i.e. the enantiomer.

In order to study the chirality of a "chain of tetrahedrons", the operator $\hat{\chi}$ has to be applied on the molecule $n$ times (one for each stereogenic centre). The molecule is completely achiral if after $\hat{\chi}^n$, the mirror image is superimposable ($p = 0$). The molecule is an enantiomer if, after $\hat{\chi}^n$, $p = n$ inversions are observed, while a diastereoisomer is obtained if, after $\hat{\chi}^n$, $p < n$ inversions are observed. In other words, the central chirality of a molecule can be expressed with respect to the base of the eigenvectors $\varphi$ (corresponding to $\lambda = 1$) and $\bar{\varphi}$ (corresponding to $\lambda = -1$). Any molecule with *n* stereogenic centres is the linear combination of the eigenvectors $\varphi$ and $\bar{\varphi}$, *p* is the number of times in which inversions of groups occur (i.e. *p* is the sum of eigenvalues $\lambda = -1$), *n* is the number of application of $\hat{\chi}$. In this sense, the couple $\chi \equiv \{n, p\}$ assigns the central chirality of a given molecule and it is a selection rule which describes achiral molecules, enantiomers and diastereoisomers. As an example, the chirality of the degenerate case *meso*-tartaric acid (Fig. 5), can be reduced to this rule. In this case, three groups of the two tetrahedrons are identical and the fourth group is the other stereogenic carbon centre. As it can be seen from the figure, by a superimposition, the molecule and its mirror image are the same ($p = 0$). The structure is achiral.

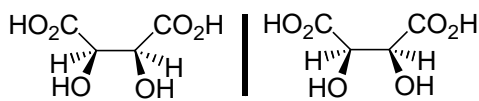

**Fig. 5.** Mirror structures of *meso*-tartaric acid

## CONCLUSIONS

In this contribution, we outlined a description of central molecular chirality of tetrahedral molecules by a geometrical approach based on complex numbers. It is essential the fact that the molecule is decomposable in tetrahedrons. An index of chirality $\chi$ has been defined as a function of *n* (number of stereogenic centres in the molecule) and *p* (number of permutations observed under rotations and superimposition to the mirror image). It is worth stressing again that this rule holds for simply connected chains of tetrahedrons, but it could be eventually generalized also for other forms of chirality, e.g. axial chirality. In the above discussion, the algebraic structure of our approach emerged. It is interesting to note that the operator $\hat{\chi}$ acts in a way similar to the Dirac matrices.[14]